\documentclass[twocolumn,showpacs,preprintnumbers,amsmath,amssymb]{revtex4}

\usepackage{graphicx}
\usepackage{dcolumn}
\usepackage{bm}
\usepackage{color}

\begin{document}

\preprint{APS/123-QED}

\title{NMR Investigation of the iron-based superconductors \\
Ca$_4$(Mg,Ti)$_3$Fe$_2$As$_2$O$_{8-y}$ and Ca$_5$(Sc,Ti)$_4$Fe$_2$As$_2$O$_{11-y}$

}

\author{Y. Tomita$^{1}$, H. Kotegawa$^{1,4}$, Y. Tao$^{1}$, H. Tou$^{1,4}$, H. Ogino$^{2,4}$, S. Horii$^{3}$, K. Kishio$^{2,4}$, and J. Shimoyama$^{2,4}$}

\affiliation{$^{1}$Department of Physics, Kobe University, Kobe 657-8501\\
$^{2}$Department of Applied Chemistry, The University of Tokyo, Tokyo 113-8656\\
$^{3}$Department of Environmental Systems Engineering, Kochi University of Technology, Kami, Kochi 782-8502\\ 
$^{4}$JST, Transformative Research - Project on Iron Pnictides (TRIP), Chiyoda, Tokyo 102-0075\\
}

\date{\today}

\begin{abstract}
$^{75}$As and $^{45}$Sc NMR measurements unravel the electronic state for Fe-based superconductors with perovskite-type blocking layers Ca$_4$(Mg,Ti)$_3$Fe$_2$As$_2$O$_{8-y}$ ($T_c^{onset}=47$ K) and Ca$_5$(Sc,Ti)$_4$Fe$_2$As$_2$O$_{11-y}$ ($T_c^{onset}=41$ K).
In Ca$_5$(Sc,Ti)$_4$Fe$_2$As$_2$O$_{11-y}$, the nuclear spin relaxation rate $1/T_1$ shows pseudogap behavior below $\sim80$ K, suggesting that the electronic state is similar to that of LaFeAs(O,F) system with moderate electron doping.
The presence of the pseudogap behavior gives an interpretation that the hole-like band (so-called $\gamma$ pocket) is located just below the Fermi level from the analogy to LaFeAs(O,F) system and the disappearance of the $\gamma$ pocket yields the suppression of the low-energy spin fluctuations.  On the other hand, in Ca$_4$(Mg,Ti)$_3$Fe$_2$As$_2$O$_{8-y}$ satisfying the structural optimal condition for higher $T_c$ among the perovskite systems, the extrinsic contribution, which presumably originates in the Ti moment, is observed in $1/T_1T$; however, the moderate temperature dependence of $1/T_1T$ appears by its suppression under high magnetic field.
In both systems, the high $T_c$ of $\sim40$ K is realized in the absence of the strong development of the low-energy spin fluctuations.
The present results reveal that the structural optimization does not induce the strong development of the low-energy spin fluctuations.
If we consider that superconductivity is mediated by spin fluctuations, the structural optimization is conjectured to provide a benefit to the development of the high-energy spin fluctuations irrespective to the low-energy part.

\end{abstract}

\maketitle

\section{Introduction}

A perovskite-type superconductor with a large two-dimensional separation of FeAs layers is one of the key examples to clarify the mechanism of high $T_c$ among Fe-based superconductors.
The maximum $T_c$ in the perovskite-type systems has been marked at $T_c^{onset}\sim47$ K for Ca$_4$(Mg,Ti)$_3$Fe$_2$As$_2$O$_{8-y}$ or $T_c^{onset}\sim46$ K for pressurized Sr$_2$VFeAsO$_3$.\cite{Ogino_47K,Kotegawa1}
The perovskite systems have a tendency that the pnictogen height is longer and the As-Fe-As bond angle is sharper compared with 1111 systems possessing the same $a$-axis length,\cite{Ogino_structure0} and the systems have many varieties of the blocking layers.
$T_c$ of over 40 K has been achieved only in 1111 systems and perovskite-type systems, thereby implying the importance of the structural feature for achieving higher values of $T_c$.

The perovskite systems are characterized by two types of composition: $A_{n+2}$$M_n$Fe$_2$As$_2$O$_{3n}$ or $A_{n+1}$$M_n$Fe$_2$As$_2$O$_{3n-1}$ ($A=$ Sr, Ca, $M=$ Sc, Mg, Ti, V, Al).\cite{Ogino_P,Zhu,Zhu2,Ogino_345,Ogino_47K,Sato,Shirage,Ogino_AlTi}
Figure 1 shows the crystal structures of the latter type, Ca$_{n+1}$$M_n$Fe$_2$As$_2$O$_{3n-1}$ ($n=3$ and $4$).
The thick blocking layer composed of Ca$_{n+1}$$M_n$O$_{3n-1}$ expands the distance between FeAs layers along the $c$ axis, and this layer distance ranges from $\sim 15 to 25$ \AA.
These values are much larger than those of other Fe-based superconductors such as FeSe (5.5 \AA) and SmFeAs(O,F) (8.5 \AA).\cite{Hsu,Chen}
It is an intriguing issue whether AF spin fluctuations are developing in the perovskite system with $T_c$ of over 40 K.

\begin{figure}[b]
\centering
\includegraphics[width=0.65\linewidth]{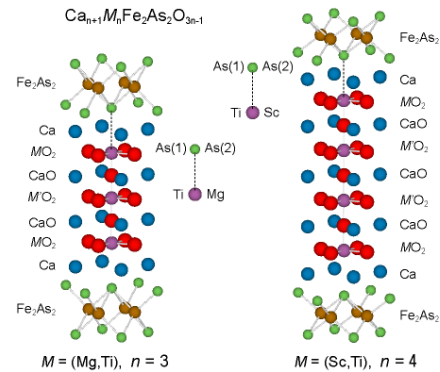}
\caption[]{(color online) The crystal structures of Ca$_4$(Mg,Ti)$_3$Fe$_2$As$_2$O$_{8}$ and Ca$_5$(Sc,Ti)$_4$Fe$_2$As$_2$O$_{11}$. The $M$ site is occupied by either Mg or Ti (Sc or Ti), thereby resulting in two inequivalent As sites. The outer $M$ site and the inner $M'$ site are crystallographically inequivalent. The $M'$ site shows nearly cubic symmetry.
}
\end{figure}

NMR is an effective tool to investigate spin fluctuations.
From the point of view of NMR measurements, close relationships between the nuclear spin relaxation rate $1/T_1$ and $T_c$ can be observed in some systems such as FeSe, pressurized SrFe$_2$As$_2$, Ba(Fe,Co)$_2$As$_2$, and BaFe$_2$(As,P)$_2$,\cite{Imai,Masaki,Kitagawa,Ning,Nakai} giving a strong suggestion for a contribution of spin fluctuations to superconductivity.
In LaFeAs(O,F), $1/T_1T$ strongly depends on the carrier doping, while the doping-dependent variation in $T_c$ seems to depend on the samples.\cite{Nakai3,Oka}
A recent study has reported the relationship between AF fluctuations and $T_c$ in La1111 systems.\cite{Oka}
However, NMR measurements for 40-50 K class high-$T_c$ Fe-based systems have not thus far been satisfactorily performed, because the large localized moment of rare-earth ions such as Sm or Nd in 1111 systems obstructs the evaluation of the intrinsic magnetic character originating in Fe moments \cite{Jeglic,Yamashita,Prando}.

\begin{table}
 \begin{center}
  \begin{tabular}{lccccc}
       & $T_c^{onset}$ (K) & $\Delta T_{c,max}^{onset}$ (K)  &  $a$ (\AA)  & Ref.  \\
   \hline
          &    &   &                  &        \\
Sr$_4$(Mg,Ti)$_2$Fe$_2$As$_2$O$_{6-y}$ & 36.7 &  6.0   &  3.936  & \cite{Sato} \\
Sr$_4$V$_2$Fe$_2$As$_2$O$_{6-y}$    & 37.2 &  9.6 & 3.930 &  \cite{Zhu} \\
Ca$_4$(Sc,Ti)$_3$Fe$_2$As$_2$O$_{8-y}$  & 34.8 & 3.1  & 3.922 & \cite{Ogino_345} \\
Ca$_5$(Sc,Ti)$_4$Fe$_2$As$_2$O$_{11-y}$  & 41.3 & 1.4  & 3.902 & \cite{Ogino_345} \\
Ca$_6$(Sc,Ti)$_5$Fe$_2$As$_2$O$_{14-y}$  & 43.3 & 0.6  &  3.884 & \cite{Ogino_345} \\
Ca$_4$(Mg,Ti)$_3$Fe$_2$As$_2$O$_{8-y}$  & 47.3 & $\sim$0.5 & 3.877 & \cite{Ogino_47K} \\
Ca$_5$(Mg,Ti)$_4$Fe$_2$As$_2$O$_{11-y}$  & 43.0 & 0 & 3.864  & \cite{Ogino_structure} \\
Ca$_8$(Mg,Ti)$_6$Fe$_2$As$_2$O$_{18-y}$  & 41.7  & 0 & 3.86  & \cite{Ogino_structure}  \\
Ca$_6$(Al,Ti)$_4$Fe$_2$As$_2$O$_{12-y}$  & 39.0  & - & 3.815 & \cite{Ogino_AlTi}  \\
Ca$_4$Al$_2$Fe$_2$As$_2$O$_{6-y}$  & 28.3  & - & 3.713 & \cite{Shirage}  \\
                               &    &         &           &       \\ 
   \hline
  \end{tabular}
    \caption{$T_c^{onset}$ at ambient pressure, the maximum $\Delta T_c^{onset}$ under pressure, and the $a$-axis length for several perovskite systems.  As for the pressure effect, two compounds are added from Ref.\cite{Kotegawa2}, which will be published elsewhere.
    }
     \end{center}
\end{table}

\begin{figure}[htb]
\centering
\includegraphics[width=0.75\linewidth]{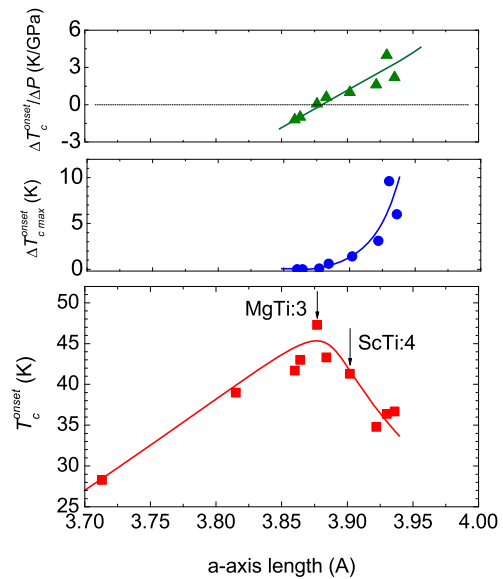}
\caption[]{(color online) The a-axis length vs. $T_c^{onset}$, the increase in $T_c$ under pressure $\Delta T_{c max}^{onset}$, and the initial slope of pressure variation of $T_c^{onset}$ up to 1 GPa for several perovskite systems. The relationship between $a$-axis length and $T_c^{onset}$ has been already summarized in Ref.\cite{Ogino_structure}. The $a$-axis length for MgTi:3 with highest $T_c$ is threshold, above which $T_c$ increases under pressure, indicating that the structural optimization is achieved for MgTi:3 among the perovskite systems. Each curve and line is a guide to the eye.
}
\end{figure}

In this paper, we report the NMR results of Ca$_4$(Mg,Ti)$_3$Fe$_2$As$_2$O$_{8-y}$ (denoted by MgTi:3 hereafter) and Ca$_5$(Sc,Ti)$_4$Fe$_2$As$_2$O$_{11-y}$ (ScTi:4).
The accurate local structures for both compounds are not determined; however MgTi:3 with $T_c^{onset}=47$ K is considered to satisfy the structural optimal conditions for higher $T_c$ among perovskite systems from systematic comparisons of the $a$-axis length and pressure effects on $T_c$.\cite{Ogino_structure,Kotegawa2}
They are summarized in Table.~I and Fig.~2. 
The clear relationship can be seen between the $a$-axis length and $T_c^{onset}$ determined by the resistivity measurement, and between the $a$-axis length and the pressure variation of $T_c$. 
$T_c^{onset}=47$ K in MgTi:3 is located at a peak position, and $T_c$ increases under pressure only when $a$ is longer than that of MgTi:3.
This tendency strongly suggests that the structural optimization is achieved in MgTi:3 among the perovskite-type systems.
The relationship between $T_c$ and the $a$-axis length also can be seen in 1111 systems,\cite{Ren} where La1111 has a longer $a$-axis length compared with Sm or Nd1111 systems with the highest $T_c$.
If two-dimensionality in the perovskite-type structure does not influence the electronic state significantly, it is briefly conjectured that the electronic state of MgTi:3 is close to Sm or Nd systems, and ScTi:4 is located between them and the La system.

\section{Experimental Procedure}

Polycrystalline samples were synthesized via solid-state reactions.\cite{Ogino_47K,Ogino_345}
The MgTi:3 exhibits $T_c^{onset}=47$ K in a resistivity measurement, which is the highest value among perovskite systems.\cite{Ogino_47K}
The presence of a clear diamagnetism is confirmed below 43 K.
The $T_c^{onset}$ of ScTi:4 is 41 K as per a resistivity measurement, and diamagnetism is observed below 36 K.\cite{Ogino_345}
The nominal composition of each compounds is given as Ca$_4$(Mg$_{0.25}$Ti$_{0.75}$)$_3$Fe$_2$As$_2$O$_{8-y}$ and Ca$_5$(Sc$_{0.5}$Ti$_{0.5}$)$_4$Fe$_2$As$_2$O$_{11-y}$, however the doping level is unclear due to a difficulty of the quantitative analysis of the stoichiometric ratio.
The samples were crushed into powder for NMR measurements.
The NMR measurements for $^{75}$As and $^{45}$Sc were performed using a conventional spin echo method in the range of a magnetic field of up to 15 T.
The $^{75}$As has a nuclear spin $I=3/2$ and a gyromagnetic ratio $\gamma_N = 7.292$ MHz/T, and $^{45}$Sc has $I=7/2$ and $\gamma_N = 10.343$ MHz/T.

\section{Experimental results and discussions}

\subsection{NMR spectrum analysis}

\begin{figure}[b]
\centering
\includegraphics[width=0.55\linewidth]{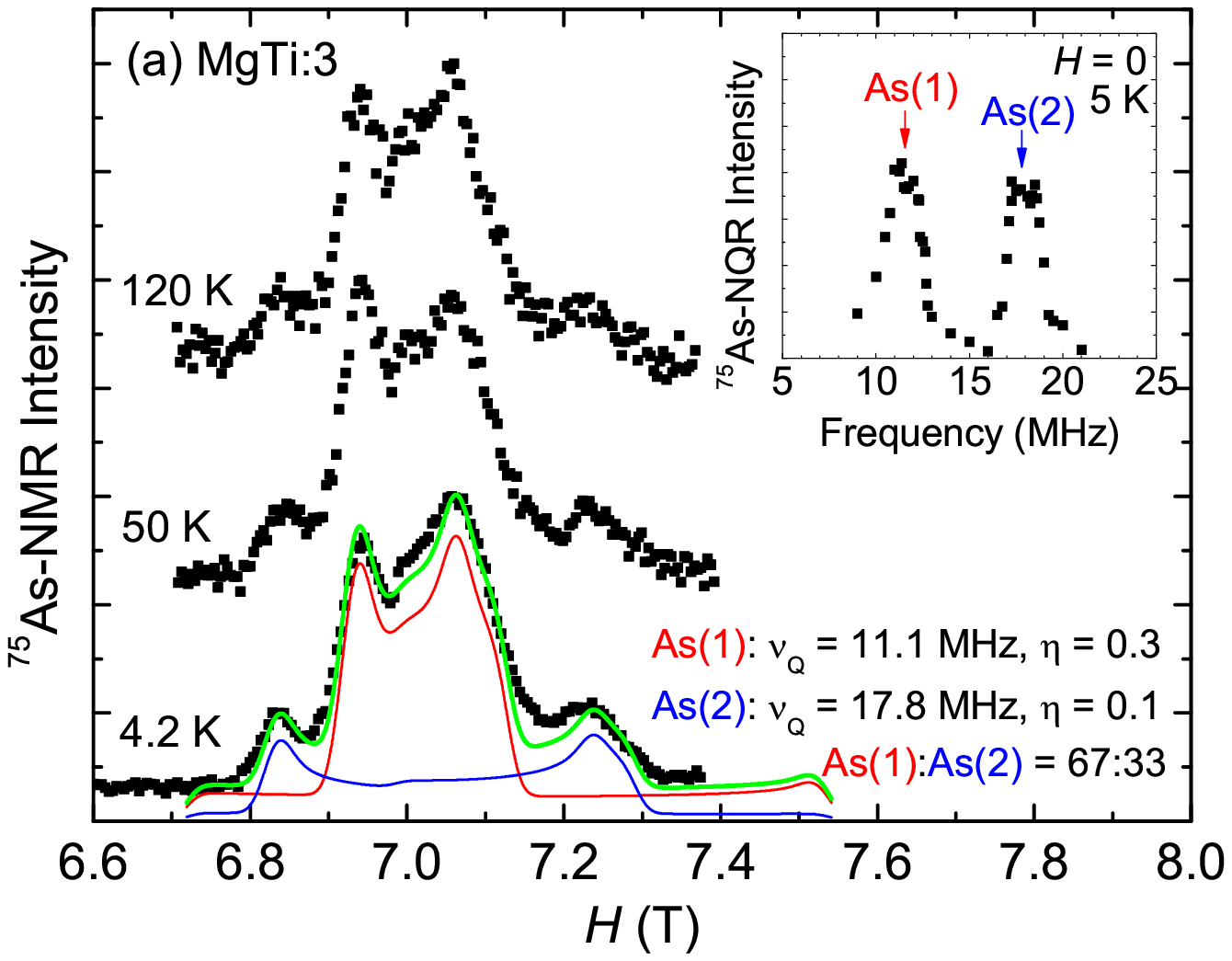}
\includegraphics[width=0.55\linewidth]{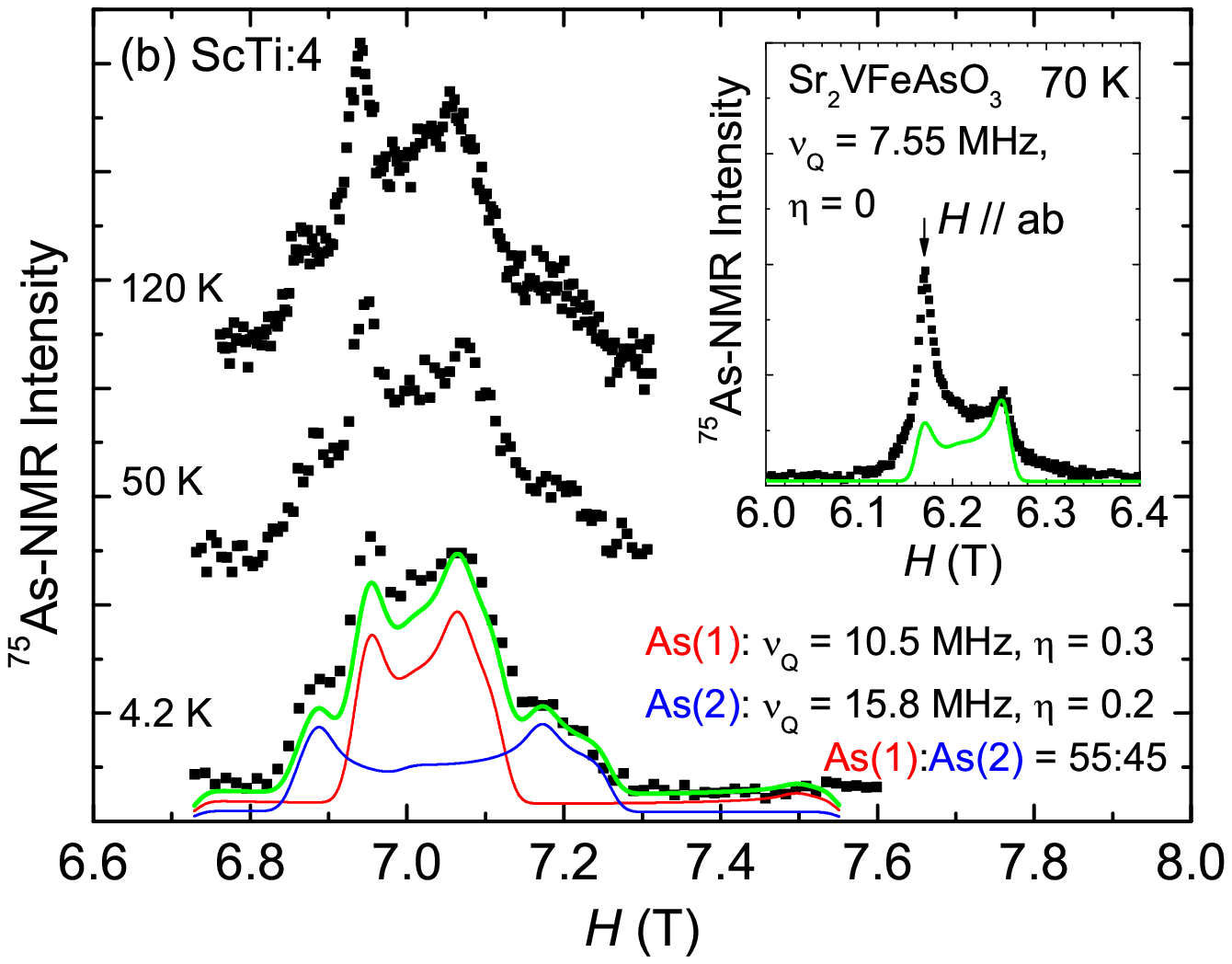}
\caption[]{(color online) (a) $^{75}$As-NMR spectrum for MgTi:3 (Ca$_4$(Mg$_{0.25}$Ti$_{0.75}$)$_3$Fe$_2$As$_2$O$_{8-y}$) measured at $\omega_0 = 51.044$ MHz. The spectrum is well reproduced by taking account of the two inequivalent As sites with each parameter shown in the figure; the inset shows the NQR spectrum. (b) $^{75}$As-NMR spectrum for ScTi:4 (Ca$_5$(Sc$_{0.5}$Ti$_{0.5}$)$_4$Fe$_2$As$_2$O$_{11-y}$). The spectrum is similarly reproduced. The $\nu_Q$ value for the As(1) site is in the range of values between those of MgTi:3 and ScTi:4, thereby suggesting that this value is attributed to the As site next to the Ti ion. The inset shows the $^{75}$As-spectrum for the stoichiometric Sr$_2$VFeAsO$_3$. The spectrum is a typical powder pattern with $\eta=0$ oriented partially along $H \parallel ab$. 
}
\end{figure}

Figures 3(a) and 3(b) show the central transitions ($-1/2 \leftrightarrow 1/2$) of the $^{75}$As spectrum measured at $\sim$7 T for MgTi:3 and ScTi:4.
Both spectra are composed of four peaks and they do not represent the typical powder patterns for FeAs superconductors which are composed of two peaks, as shown for the case of Sr$_2$VFeAsO$_3$ in the inset of Fig.~3(b), where the crystal is oriented partially along $H \parallel ab$.
The present spectra are rather similar to that of the perovskite system Ca$_6$(Al,Ti)$_4$Fe$_2$As$_2$O$_{12-y}$.\cite{Nakano}
Nakano {\it et al.} have shown that the spectrum can be reproduced by taking into account the asymmetry parameter of the electric field gradient $\eta$.\cite{Nakano}
However, the four distinct peaks shown in Figs.~3 cannot be reproduced only by a powder pattern with $\eta$ for a single As site.
The inset in Fig. 3(a) shows the $^{75}$As-NQR spectrum ($\pm1/2 \leftrightarrow \pm3/2$ transition) measured at zero field for MgTi:3, and it reveals the existence of two inequivalent As sites.
This inequivalency can be understood crystallographically, because the neighboring $M$ site is occupied by either Mg or Ti as shown in Fig.~1, and the quadrupole frequency ($\nu_Q$) strongly depends on the surrounding ions.
For instance, the As site of Sr$_2$VFeAsO$_3$ ($A=$ Sr, $M=$ V) has $\nu_Q \sim 7.55$ MHz, and that of Ca$_2$AlFeAsO$_{3-y}$ [Ca$_4$Al$_2$Fe$_2$As$_2$O$_{6-y}$] ($A=$ Ca, $M=$ Al) has $\nu_Q \sim 22.6$ MHz.\cite{Kotegawa2,Kinouchi}
In actuality, the NMR spectrum for MgTi:3 is well reproduced by a powder pattern taking into account two As sites with $\nu_Q=11.1$ MHz, $\eta=0.3$ [As(1)] and $\nu_Q=17.8$ MHz, $\eta=0.1$ [As(2)]; these values nearly correspond to the observed NQR frequencies.
In this spectral calculation, the distribution of $\nu_Q$ is also taken into account.
The asymmetry parameter $\eta$ is conjectured to be induced by the breaking of the local four-fold symmetry at the As site due to the random distribution of the second-nearest-neighboring and more distant neighboring $M$ site, because $\eta$ is almost zero in the stoichiometric system Sr$_2$VFeAsO$_3$.\cite{Kotegawa2}
In this sense, $\eta$ is expected to distribute in the crystal and the obtained fitting parameter indicates the averaged value.
Similarly, the spectrum of ScTi:4 is also reproduced by two inequivalent As sites with $\nu_Q=10.5$ MHz, $\eta=0.3$ [As(1)] and $\nu_Q=15.8$ MHz, $\eta=0.2$ [As(2)].
From a comparison of the intensity ratio between two As sites, the As(1) site with the similar frequencies of $\sim11$ MHz for both compounds is expected to be the As site next to the Ti ion.
The shapes of the $^{75}$As-NMR spectra for both compounds are nearly unchanged with variation in temperature.
However, we could not evaluate the accurate temperature dependence of the Knight shift because of the error in the estimation of $\nu_Q$ and $\eta$, which also depend on temperature.

\begin{figure}[b]
\centering
\includegraphics[width=0.6\linewidth]{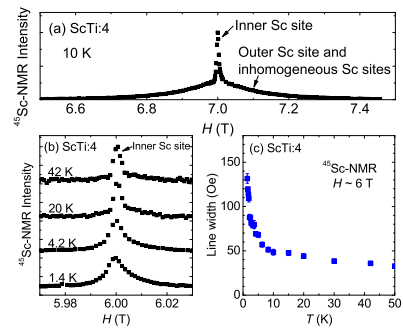}
\caption[]{(color online) (a) $^{45}$Sc-NMR spectrum at 10 K, which indicates two Sc sites. (b) $^{45}$Sc-NMR spectrum for the inner Sc site at several temperatures. The spectrum clearly broadens at low temperatures. (c) Temperature dependence of line-width of $^{45}$Sc-NMR spectrum.
}
\end{figure}

Figure 4(a) shows the $^{45}$Sc-NMR spectrum for ScTi:4 measured at $\sim7$ T. 
The $^{45}$Sc-NMR spectrum is composed of a sharp peak observed at around zero Knight shift ($H_0 \sim 7$ T) along with broad tails.
Since the nuclear spin of $^{45}$Sc nuclei is $I=7/2$, the broad tails can be interpreted as the quadrupole satellite lines smeared out by the random $\nu_Q$.
On the other hand, a sharp peak indicates the Sc site with $\nu_Q \sim 0$, which is in a high symmetry close to a cubic one.
These two different signals are attributed to the inequivalent Sc sites shown in Fig.~1.
The inner Sc site denoted by $M'$ is located within a nearly cubic symmetry resulting in $\nu_Q\sim 0$. 
The outer Sc site denoted by $M$ shows lower symmetry, so that $\nu_Q \ne 0$.
Figure 4(b) shows the spectra for the inner Sc site at several temperatures.
It is observed that the spectrum broadens at low temperatures.
Figure 4(c) shows the temperature dependence of the spectral width.
The width suddenly increases below $\sim8$ K, thereby indicating the establishing of the internal magnetic field.
This behavior is similar to that of the $^{27}$Al spectrum in Ca$_6$(Al,Ti)$_4$Fe$_2$As$_2$O$_{12-y}$, \cite{Nakano} where the $^{75}$As-spectrum also broadens at low temperatures due to the larger internal field when compared with those in the present cases.
The magnitude of the internal field is smaller in ScTi:4 by a factor of $1/4-1/5$ when compared with that of Ca$_6$(Al,Ti)$_4$Fe$_2$As$_2$O$_{12-y}$ at the Sc/Al site.

\subsection{Spin dynamics in perovskite-type systems}

\begin{figure}[b]
\centering
\includegraphics[width=0.6\linewidth]{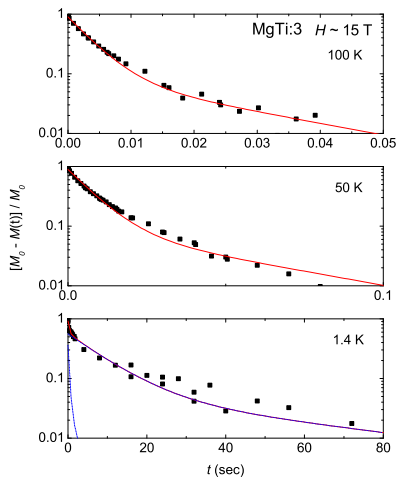}
\caption[]{(color online) The recovery curves for $^{75}$As-NMR for MgTi:3 measured at $\sim15$ T. In the normal state, the recovery curves are well-fitted by the theoretical curve. 
In the SC state, the short component appears due to the vortex core and the influence from the fluctuations of the small localized moment.
}
\end{figure}

Figure 5 shows the recovery curves at the As site for MgTi:3 to evaluate the nuclear spin - lattice relaxation time $T_1$.
The $T_1$ was measured at the As(1) site for $H \parallel ab$.
In the normal state, the recovery curve is well fitted by the theoretical curve for the central transition of $I=3/2$, thereby indicating that the magnetic fluctuation is dominated by the specially homogeneous one.
In the SC state, since the recovery curves cannot be fitted by a single component of $T_1$, as shown in the figure, we fitted the data by assuming two components.
The short component is attributed to the electronic state in the vortex core and the region strongly affected by the magnetic moment inducing the static internal field at low temperatures, which presumably originates in Ti 3$d$ electrons.
We used the long component to discuss the electronic state in the SC state, although the contribution from the extrinsic fluctuation cannot be excluded sufficiently.

\begin{figure}[htb]
\centering
\includegraphics[width=0.6\linewidth]{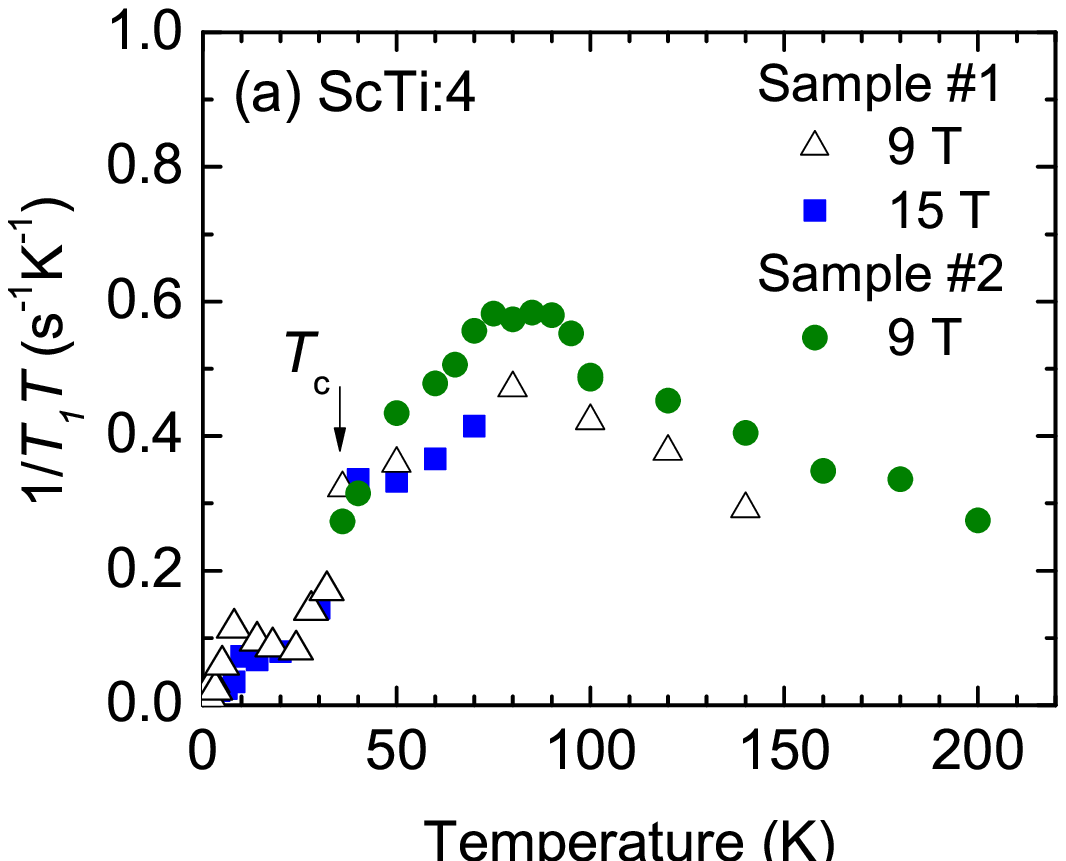}
\includegraphics[width=0.6\linewidth]{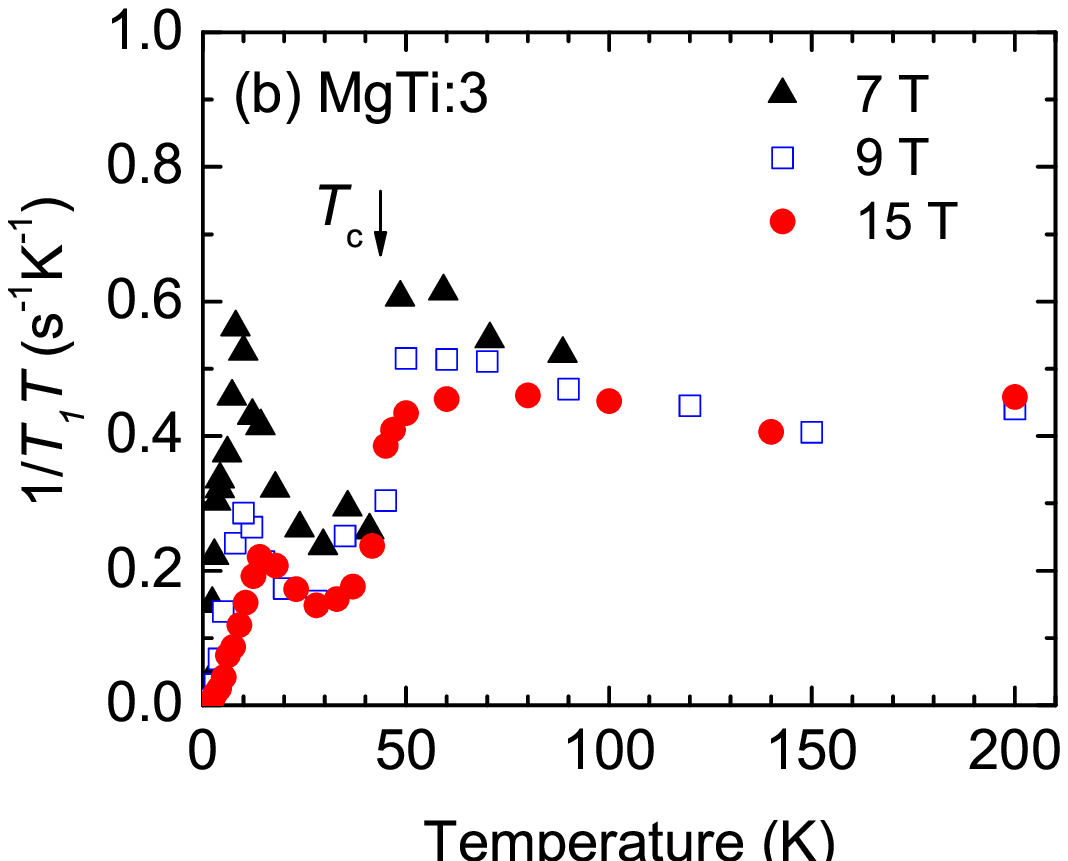}
\caption[]{(color online) Temperature dependences of $1/T_1T$ for (a) ScTi:4 and (b) MgTi:3. In the normal state for ScTi:4, pseudogap behavior is observed below $\sim80$ K irrespective of samples, which were made by a same procedure but the carrier density is conjectured to be slightly different. In MgTi:3, the moderate temperature dependence is seen at high magnetic field, where the extrinsic fluctuation is suppressed. In the SC state, the peaks appear at around 10 K due to the critical slowing down of some localized moment.  
}
\end{figure}

Figure 6 shows the temperature dependence of $1/T_1T$ for (a) ScTi:4 and (b) MgTi:3 measured at different magnetic fields.
In both samples, the clear drops in $1/T_1T$ below $T_c$ are observed in the absence of a coherence peak, thereby confirming that the superconductivity is of bulk nature.
In the SC state, the peaks in $1/T_1T$ appear at around 10 K, and it is more pronounced for MgTi:3.
Taking into account the broadening of the $^{45}$Sc spectrum for ScTi:4 below $\sim8$ K, the peaks indicate a critical slowing down of some magnetic moment.
As clearly seen in MgTi:3, the temperature of this peak increases under high magnetic field, thereby indicating that this magnetic ordering is not a typical AF one.

In the normal state for ScTi:4, Curie-Weiss like behavior at high temperatures and pseudogap behavior below $\sim80$ K appear irrespective of samples.
The peak of $\sim80$ K does not originate in the magnetic ordering because of the absence of the spectral broadening.
The pseudogap behavior in the similar temperature range has been seen for La1111 systems in the lightly or optimally doped region without the magnetic ordering: $x=0.07$ ($T_c=22.5$ K) or $x=0.11$ ($T_c=21$ K).\cite{Nakai3,Kobayashi}
According to band calculations, the pseudogap behavior indicates the presence of the high density of states just below the Fermi level,\cite{Ikeda} which mainly originates in the band at $(\pi,\pi)$ in unfolded Brillouin zone (so-called $\gamma$ pocket if it crosses the Fermi level).\cite{Kuroki2}
If the $\gamma$ pocket takes part in the Fermi surface, the nesting between the $\gamma$ pocket and the electron-like pockets (so-called $\beta$) yields the low-energy spin fluctuations which strongly develops towards low temperatures; however if it is absent, the pseudogap appears and the moderates temperature dependence of $1/T_1T$ is realized.\cite{Ikeda,Kuroki}
In this context, $1/T_1T$, which proves low-energy spin fluctuations of $\sim10^1 -10^2$ MHz, is governed by mainly the presence or absence of the $\gamma$ pocket, and the Fermi surface of ScTi:4 is conjectured to be almost lacking the $\gamma$ pocket.

In MgTi:3, on the other hand, $1/T_1T$ shows the field dependence below $\sim100$ K.
Generally the spin fluctuation originating in the Fe moment is robust against magnetic field.\cite{Nohara} 
The observed increase in $1/T_1T$ depending on magnetic field is attributed in other extrinsic moment interacts weakly, which is probably a Ti-localized moment.
$T_1T \sim const.$ behavior at 15 T, which appears as a consequence of the suppression of the extrinsic $1/T_1T$, suggests that the intrinsic $1/T_1T$ is temperature-independent or decreases with decreasing temperature.
It is difficult to mention whether the pseudogap behavior is present or not because a high field of 15 T may not be enough to suppress the extrinsic relaxation completely.
If we consider that the $1/T_1T$ at higher field is inherent in the spin fluctuation originating in the Fe moment, it is clear that the strong magnetic criticality is absent in MgTi:3.
It is conjectured that the $\gamma$ pocket does not appear obviously in the Fermi surface of MgTi:3 as well as other electron-doped systems.\cite{Ikeda}
It should be noted that the absence of the strong magnetic criticality does not mean the absence of the AF spin fluctuations.
It is interpreted that the low-energy part of the spin fluctuations is temperature independent.\cite{Ikeda}
Even in the La1111 system with moderate temperature dependence of $1/T_1T$, the anisotropy of $1/T_1T$ has been understood to originate in the stripe-type AF spin fluctuation.\cite{SKitagawa}
In addition to the present results, the moderate temperature dependence in $1/T_1T$ has been also observed in Sr$_4$(Mg,Ti)$_2$Fe$_2$As$_2$O$_{6-y}$ whose $a$-axis length is relatively longer.\cite{Yamamoto} 
The strong enhancement of $1/T_1T$ has been reported in Ca$_4$Al$_2$Fe$_2$As$_2$O$_{6-y}$ with the quite shorter $a$-axis length.\cite{Kinouchi}
In the longer $a$-axis regime, to which the La1111 system also belongs, there is no clear indication that the two-dimensional separation of FeAs layers in the perovskite-type structure gives a significant enhancement of the spin fluctuations, at least in the low-frequency region.

\begin{figure}[htb]
\centering
\includegraphics[width=0.6\linewidth]{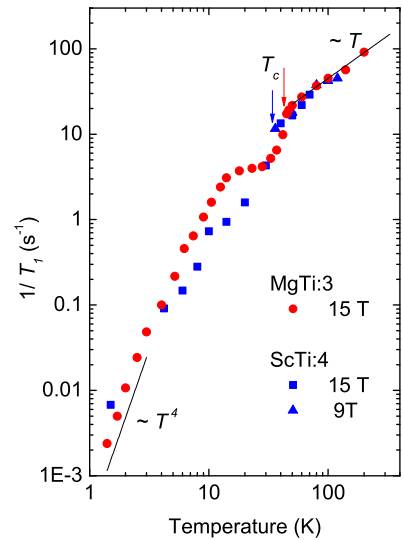}
\caption[]{(color online)  Temperature dependence of $1/T_1$ for MgTi:3 and ScTi:4. $1/T_1$ continues to decrease even at low temperatures. 
}
\end{figure}

Figure 7 shows the temperature dependencies of $1/T_1$ for MgTi:3 and ScTi:4.
The fluctuations in the localized moment make it difficult to evaluate the SC gap symmetry in these compounds.
However, particularly in MgTi:3, $1/T_1$ shows a steep decrease close to a $T^4$ relation even at the lowest temperature without exhibiting $T_1T$=const. behavior originating in the residual density of state (DOS) near the Fermi level.
The residual DOS for MgTi:3 is estimated to be less than 5\% of the DOS in the normal state.
In P-doped 122 systems which have been pointed out to have nodes in the SC gap, a large residual DOS of more than 30\% is observed in $T_1$ measurements,\cite{Nakai2,Dulguun} and this is in strong contrast to the present case.
The SC gap symmetry of MgTi:3 is conjectured to be nodeless as in many Fe-based superconductors, but more careful investigation is required.

The present NMR results revealed that the high $T_c$ of over 40 K in MgTi:3, which is considered to satisfy the structural optimal condition as shown in Fig.~1,\cite{Ogino_structure,Kotegawa2} is realized in the moderate low-energy spin fluctuations.
The situation might be similar to that of PrFeAs(O,F) with $T_c=45$ K, where the temperature dependence of $1/T_1T$ is weak in the normal state.\cite{Matano}
These behavior are in contrast to the cases in FeSe and some 122 systems.\cite{Imai,Masaki,Kitagawa,Ning,Nakai}
In 122 systems, the higher $T_c$ is realized accompanied by quantum critical behavior, and superconductivity disappears along with the disappearance of the criticality.
These results suggest that the spin fluctuations are relevant with superconductivity and also imply that the high-energy part of the spin fluctuations develops along with the low-energy part in these systems.
In the moderately doped LaFeAs(O,F) and LiFeAs, superconductivity is realized in the absence of strong critical fluctuations.\cite{Nakai3,Oka,Kobayashi,Li}
The present perovskite systems possess about twofold higher $T_c$ than those of LaFeAs(O,F) and LiFeAs, however there is no clear indication that the spin fluctuation develops compared with them.
This suggests that the development of the low-energy spin fluctuation is not important for achieving high $T_c$.
If we consider that superconductivity is mediated by spin fluctuations, it is naturally considered that only the high-energy part develops in the present perovskite-type systems irrespective of the low-energy part, compared with La1111 systems.
This may be a common feature in the Fe-based superconductors being located near the pseudogap regime, in which $\gamma$ pocket is absent or small.
For example, in the calculation for the electron-doped pseudogap regime, only high-energy spin fluctuations develop with decreasing temperature in contrast to the hole-doped regime.\cite{Ikeda}
Based on the theoretical suggestion taking into account of the nesting properties,\cite{Usui} a best condition for high $T_c$ is achieved in the presence of the small $\gamma$ pocket.
In this condition, the low-energy spin fluctuation is not strongly enhanced owing to the small $\gamma$ pocket,\cite{Kuroki} giving no contradiction to the present results.

\section{Conclusion}

In summary, we performed NMR measurements for Ca$_4$(Mg,Ti)$_3$Fe$_2$As$_2$O$_{8-y}$ and Ca$_5$(Sc,Ti)$_4$Fe$_2$As$_2$O$_{11-y}$ which have high $T_c$ values.
The $^{75}$As- and $^{45}$Sc-spectral shapes were well explained by taking into account the inequivalency of the As and Sc sites. 
The magnetic ordering of some extrinsic moment, which is conjectured to be the Ti moment, occurs at low temperatures, and its contribution to $T_1$ obstructs a minute investigation of the SC gap symmetry.
The small $1/T_1T$ at the lowest temperature in Ca$_4$(Mg,Ti)$_3$Fe$_2$As$_2$O$_{8-y}$ indicates the upper limit of the residual DOS in the SC state is $\sim5$ \% of that in the normal state.
The gap symmetry is likely to be nodeless, however careful measurement is needed for elucidating it.
In Ca$_5$(Sc,Ti)$_4$Fe$_2$As$_2$O$_{11-y}$, the pseudogap behavior in $1/T_1T$ suggests that the hole-like band, which is related to so-called $\gamma$ pocket, is located just below the Fermi surface as well as the moderately doped La1111 systems.
The disappearance of the pocket suppresses the development of the low-energy spin fluctuations significantly, however relatively high $T_c$ of $\sim40$ K is realized.
The high $T_c$ of over 40 K in Ca$_4$(Mg,Ti)$_3$Fe$_2$As$_2$O$_{8-y}$ is also realized without the clear indication of the strong low-energy spin fluctuations.
The observed temperature dependencies of $1/T_1T$ in the high-$T_c$ perovskite-type systems are almost similar to those of La1111 systems in spite of the twofold higher $T_c$.
The present results suggest that the structural optimization for high $T_c$ does not induce the strong development of low-energy spin fluctuations, and thus it is rather conjectured to be beneficial to the development of the high-energy spin fluctuations.

\section*{Acknowledgement}

The authors thank Kazuhiko Kuroki for helpful discussions.
This work has been partially supported by Grants-in-Aid for Scientific Research (Nos. 22740231, 22013011 and 20102005 (Innovative Area ''Heavy Electrons'')) from the Ministry of Education, Culture, Sports, Science, and Technology (MEXT) of Japan.


\begin{references}













\bibitem{Ogino_47K}
H. Ogino, Y. Shimizu, K. Ushiyama, N. Kawaguchi, K. Kishio, and J. Shimoyama, Applied Physics Express {\bf 3}, 063103 (2010).


\bibitem{Kotegawa1}
H. Kotegawa, T. Kawazoe, H. Tou, K. Murata, H. Ogino, K. Kishio, and J. Shimoyama, J. Phys. Soc. Jpn. {\bf 78}, 123707 (2009).


\bibitem{Ogino_structure0}
H. Ogino, S. Sato, Y. Matsumura, N. Kawaguchi, K. Ushiyama, Y. Katsura, S. Horii, K. Kishio, and J. Shimoyama, Physica C {\bf 470}, S280 (2010).





\bibitem{Ogino_P}
H. Ogino, Y. Matsumura, Y. Katsura, K. Ushiyama, S. Horii, K. Kishio, and J. Shimoyama: Supercond. Sci. Technol. {\bf 22}, 075008 (2009).


\bibitem{Zhu}
X. Zhu, F. Han, G. Mu, P. Cheng, B. Shen, B. Zeng, and H.-H. Wen, Phy. Rev. B {\bf 79}, 220512(R) (2009).


\bibitem{Zhu2}
X. Zhu, F. Han, G. Mu, B. Zeng, P. Cheng, B. Shen, H. -H. Wen, Phys. Rev. B {\bf 79}, 024516 (2009).


\bibitem{Sato}
S. Sato H. Ogino, N. Kawaguchi, Y. Katsura, K. Kishio, J. Shimoyama, H. Kotegawa, H. Tou: Supercond. Sci. Technol. {\bf 23} (2010) 045001. 


\bibitem{Ogino_345}
H. Ogino, S. Sato, K. Kishio, J. Shimoyama, T. Tohei, and Y. Ikuhara, Appl. Phys. Lett. {\bf 97}, 072506 (2010).









\bibitem{Shirage}
P. M. Shirage, K. Kihou, C.-H Lee, H. Kito, H. Eisaki, and A. Iyo, Appl. Phys. Lett. {\bf 97}, 172506 (2010).



\bibitem{Ogino_AlTi}
H. Ogino, K. Machida, A. Yamamoto, K. Kishio, J. Shimoyama, T. Tohei, Y. Ikuhara: Supercond. Sci. Technol. {\bf 23} (2010) 115005.




\bibitem{Hsu}
F.-C. Hsu, J.-Y. Luo, K.-W Yeh, T.-K. Chen, T.-W. Huang, P. M. Wu, Y.-C. Lee, Y.-L. Huang, Y.-Y. Chu, D.-C. Yan, and M.-K. Wu, PNAS {\bf 105}, 14262 (2008).


\bibitem{Chen}
X. H. Chen, T. Wu, G. Wu, R. H. Liu, H. Chen and D. F. Fang, Nature(London) {\bf 453}, 761 (2008).






\bibitem{Imai}
T. Imai, K. Ahilan, F. L. Ning, T. M. McQueen, and R. J. Cava, Phys. Rev. Lett. {\bf 102}, 177005 (2009).


\bibitem{Masaki}
S. Masaki, H. Kotegawa, Y. Hara, H. Tou, K. Murata, Y. Mizuguchi, and Y. Takano, J. Phys. Soc. Jpn. {\bf 78}, 063704 (2009).


\bibitem{Kitagawa}
K. Kitagawa, N. Katayama, H. Gotou, T. Yagi, K. Ohgushi, T. Matsumoto, Y. Uwatoko, and M. Takigawa, Plys. Rev. Lett. {\bf 103}, 257002 (2009).



\bibitem{Ning}
F. L. Ning, K. Ahilan, T. Imai, A. S. Sefat, M. A. McGuire, B. C. Sales, D. Mandrus, P. Cheng, B. Shen, and H.-H Wen, Phys. Rev. Lett. {\bf 104}, 037001 (2010).

\bibitem{Nakai} 
Y. Nakai, T. Iye, S. Kitagawa, K. Ishida, H. Ikeda, S. Kasahara, H. Shishido, T. Shibauchi, Y. Matsuda, and T. Terashima, Phys. Rev. Lett. {\bf 105}, 107003 (2010).



\bibitem{Nakai3}
Y. Nakai, S. Kitagawa, K. Ishida, Y. Kamihara, M. Hirano, and H. Hosono, Phys. Rev. B {\bf 79}, 212506 (2009).


\bibitem{Oka}
T. Oka, Z. Li, S. Kawasaki, G. F. Chen, N. L. Wang, and G. -q. Zheng, Phys. Rev. Lett. {\bf 108}, 047001 (2012).



\bibitem{Jeglic}
P. Jegli$\check{\rm c}$, J.-W. G. Bos, A. Zorko, M. Brunelli, K. Koch, H. Rosner, S. Margadonna, and D. Ar$\hat{\rm c}$on, Phys. Rev. B {\bf 79}, 094515 (2009).


\bibitem{Yamashita}
H. Yamashita, M. Yashima, H. Mukuda, Y. Kitaoka, P. M. Shirage, and A. Iyo, Physica C {\bf 470}, S375 (2010).


\bibitem{Prando}
G. Prando, P. Carretta, A. Rigamonti, S. Sanna, A. Palenzona, M. Putti, and M. Tropeano, Phys. Rev. B {\bf 81}, 100508(R) (2010)




\bibitem{Ogino_structure}
H. Ogino, Y. Shimizu, N. Kawaguchi, K. Kishio, J. Shimoyama, T. Tohei and
Y. Ikuhara, Supercond. Sci. Technol. {\bf 24} (2011) 085020.



\bibitem{Kotegawa2}
H. Kotegawa, Y. Tao, H. Tou, H. Ogino, S. Horii, K. Kishio, and J. Shimoyama, J. Phys. Soc. Jpn. {\bf 80}, 014712 (2011).


\bibitem{Ren}
Z.-A. Ren, G.-C. Che, X.-L. Dong, J. Yang, W. Lu, W. Yi, X.-L. Shen, Z.-C. Li, L.-L. Sun, F. Zhou and Z.-X. Zhao, EPL, {\bf 83} (2008) 17002.



\bibitem{Nakano}
T. Nakano, N. Fujiwara, S. Tsutsumi, H. Ogino, K. Kishio, and J. Shimoyama, Phys. Rev. B {\bf 84}, 060502(R) (2011).



\bibitem{Kinouchi}
H. Kinouchi, H. Mukuda, M. Yashima, Y. Kitaoka, P. M. Shirage, H. Eisaki, and A. Iyo, Phys. Rev. Lett. {\bf 107}, 047002 (2011).





\bibitem{Kobayashi}
Y. Kobayashi, E. Satomi, S. C. Lee, and M. Sato, J. Phys. Soc. Jpn. {\bf 79}, 093709 (2010).

\bibitem{Ikeda}
H. Ikeda, J. Phys. Soc. Jpn. {\bf 77}, 123707 (2008).

\bibitem{Kuroki2}
K. Kuroki, H. Usui, S. Onari, R. Arita, and H. Aoki, Phys. Rev. B {\bf 79}, 224511 (2009).

\bibitem{Kuroki}
K. Kuroki, private communication.


\bibitem{Nohara}
H. Nohara, H. Kotegawa, Y. Hara, H. Tou, Y. Mizuguchi, Y. Takano, J. Phys. Soc. jpn. {\bf 80}, (2011) SA120.





\bibitem{Yamamoto}
K. Yamamoto, H. Mukuda, H. Kinouchi, M. Yashima, Y. Kitaoka, M. Yogi, S. Sato, H. Ogino, and J. Shimoyama, J. Phys. Soc. Jpn. {\bf 81} 053702 (2012).


\bibitem{SKitagawa}
S. Kitagawa, Y. Nakai, T. Iye, K. Ishida, Y. Kamihara, M. Hirano, and H. Hosono: Phys. Rev. B {\bf 81}, 212502 (2010).





\bibitem{Nakai2}
Y. Nakai, T. Iye, S. Kitagawa, K. Ishida, S. Kasahara, T. Shibauchi, Y. Matsuda, and T. Terashima, Phys. Rev. B {\bf 81}, 020503(R) (2010)


\bibitem{Dulguun}
T. Dulguun, H. Mukuda, H. Kinouchi, M. Yashima, Y. Kitaoka, T. Kobayashi, S. Miyasaka, and S. Tajima, Phys. Rev. B {\bf 85}, 144515 (2012).

\bibitem{Matano}
K. Matano, Z. A. Ren, X. L. Dong, L. L. Sun, Z. X. Zhao, and G. -q. Zheng, EPL, {\bf 83}, 57001 (2008).


\bibitem{Li}
Z. Li, Y. Ooe, X.-C. Wang, Q.-Q. Liu, C.-Q. Jin, M. Ichioka, and G.-q. Zheng, J. Phys. Soc. Jpn. {\bf 79}, 083702 (2010).


\bibitem{Usui}
H. Usui, K. Suzuki, and K. Kuroki, Supercond. Sci.Technol. {\bf 25}, (2012) 084004.


\end{references}
\end{document}